\documentclass[a4paper,11pt]{article}

\usepackage{jinstpub} 
\usepackage{multirow}
\usepackage{upgreek}
\usepackage{lineno}

\title{Microbulk Micromegas in non-flammable mixtures of argon and neon at high pressure}

\author[a,1]{F.J.~Iguaz~\note{Permanent address: Synchrotron Soleil, BP 48, Saint-Aubin, 91192 Gif-sur-Yvette, France}}
\author[a, b, c, 2]{T.~Dafni~\note{Corresponding author.}}
\author[a]{C.~Canellas}
\author[a, b, c]{J.~F.~Castel}
\author[a, b, c]{S.~Cebri\'an}
\author[a]{J.~G.~Garza}
\author[a, b, c]{I.~G.~Irastorza}
\author[a, b, c]{G.~Luz\'on}
\author[a, b, c]{H.~Mirallas}
\author[a, 3]{E.~Ruiz~Ch\'oliz~\note{Present address: ETAP - Experimentelle Teilchen- und Astroteilchen Physik, Johannes Gutenberg University Mainz, Staudingerweg 7, 55128 Mainz, Germany}}
\affiliation[a]{Departamento de F\'isica Te\'orica,
Universidad de Zaragoza, C/ Pedro Cerbuna 12, 50009, Zaragoza, Spain.}
\affiliation[b]{CAPA, Centro de Astropart\'iculas y F\'isica de Altas Energ\'ias, C/ Pedro Cerbuna 12, 50009, Zaragoza, Spain.}
\affiliation[c]{Laboratorio Subterr\'aneo de Canfranc,Paseo de los Ayerbe s/n, 22880 Canfranc Estación, Huesca, Spain.}
\emailAdd{tdafni@unizar.es}

\abstract{We report on a systematic characterization of microbulk Micromegas readouts in high-pressure Ar+1\%iC$_4$H$_{10}$ and Ne+2\%iC$_4$H$_{10}$ mixtures. Experimental data on gain, electron transmission and energy resolution are presented for a wide range of drift and amplification voltages and pressures from 1\,bar to 10\,bar for the argon mixture and from 5\,bar to 10\,bar in the neon mixture, in steps of 1\,bar. Maximum gains higher than 1.7$\times 10^3$ (1.7$\times 10^4$) in the argon (neon) mixture are measured for all pressures, without the significant decrease with pressure typically observed in other amplification structures. A competitive energy resolution at 22.1\,keV, but with a slight degradation with pressure, is observed: from 10.8\% at 1\,bar to 15.6\%~FWHM at 10\,bar in Ar+1\%iC$_4$H$_{10}$ and from 8.3\% at 5\,bar to 15.0\%~FWHM at 10\,bar in Ne+2\%iC$_4$H$_{10}$. The experimental setup, procedure and the results will be presented and discussed in detail. The work is motivated by the TREX-DM experiment, that is operating in the Laboratorio Subterráneo de Canfranc with the mentioned mixtures, although the results may be of interest for other applications of time projection chambers at high pressures.}


\begin{document}
\maketitle
\flushbottom

\section{Introduction}
\label{sec:Motivation}

Charge amplification in gas is the basis for the readout of ionization signals in conventional gas time projection chambers (TPCs). Normally optimized for operation at pressures close to atmospheric, the performance of TPC charge readouts typically decreases with pressure. This is because the effects limiting the normal evolution of the avalanche, leading to streamers and eventually discharges between the electrodes, appear earlier (i.e. at lower gains) in higher gas densities. The precursor of these effects is the production of fluorescence photons in the avalanche, able to initiate secondary avalanches at points separate from the primary avalanche location. The use of quenchers, small amounts of additive gases that are opaque to these photons, is the usual strategy to limit these effects and obtain stable operation, although with an effectiveness that may decrease at higher gas pressures. It is expected that the particular geometry of the amplification structure and the avalanche also play a role, and that different types of readouts show this effect to a different extent.

It is a general observation that gain amplification substantially decreases when going to higher pressures above atmospheric. For example, in~\cite{Bondar:2001mt} the maximum gain of a triple-GEM readout in pure argon is measured to decrease from $10^5$ at 1~bar down to $5 \times 10^2$ at 5\,bar. Among the different types of micro-pattern gas detector (MPGD) technologies, it is known that Micromegas readouts compare favourably in this regard, something that we have tentatively attributed to the higher geometrical confinement of the avalanche \cite{CALCATERRA2006444}. In a Micromegas readout all the amplification takes place in the $\sim$50-100\,$\upmu$m gap between a micromesh and the anode plane on which it is laid, suspended by insulator pillars. For example, bulk Micromegas operating in Ar+2\%iC$_4$H$_{10}$ have shown a maximum gain drop from 3$\times 10^3$ at 1.2\,bar down to 5$\times 10^2$ at 10\,bar~\cite{Iguaz:2015myh}.

\begin{figure}[t]
\centering
\includegraphics[width=0.65\textwidth]{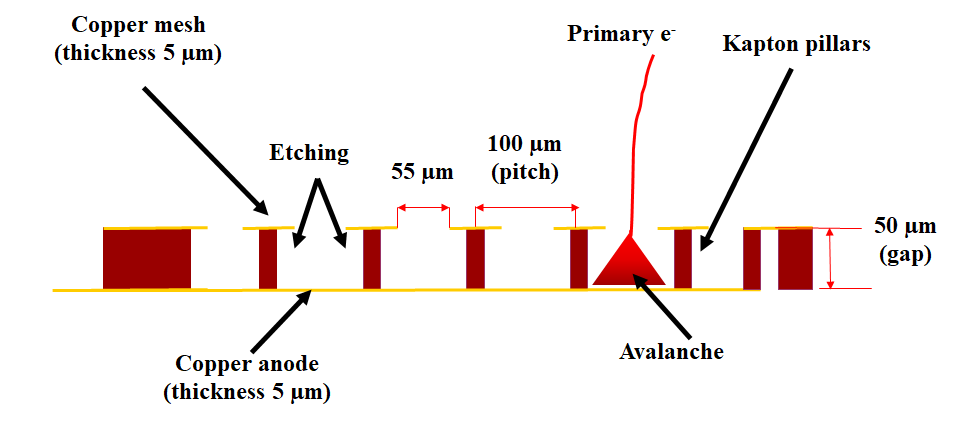}
\caption{Sketch of a microbulk Micromegas with the same characteristics (gap size, mesh hole diameter and mesh hole pitch) as the one used in these measurements.}
\label{fig:Micromegas}
\end{figure}

Microbulk Micromegas~\cite{Andriamonje:2010zz,Iguaz:2012ur} are a novel flavour of Micromegas readouts, in which the amplification geometry is obtained out of doubly copper-clad kapton foils, by means of lithography techniques and kapton etching.
The resultant readout consists of a thin copper mesh separated from the anode plane by a kapton structure (figure~\ref{fig:Micromegas}) . This technique offers some improvements over conventional Micromegas: the amplification gap is more homogeneous and the mesh is thinner, factors that reduce the avalanche fluctuations and improve the energy resolution, to as low as 11\% (10.5\%)\,FWHM at 5.9\,keV in argon-(neon-)based mixtures at 1\,bar~\cite{Iguaz:2012ur} and of 7.3\%\,FWHM at 22.1\,keV xenon-trimethylamine (TMA) at 1\,bar~\cite{Cebrian:2012sp}.

There is an important geometrical feature in microbulk detectors, due to the way they are fabricated, by etching the kapton through the holes in the copper mesh. The amplification gap is opened right below each hole in the mesh, forming a cylindrical space in the kapton of diameter slightly larger than the diameter of the corresponding mesh hole. In all the remaining area the kapton is left untouched. This is different than in bulk detectors, in which the insulator forms cylindrical pillars periodically placed along the gap. This means that the avalanche in microbulk detectors is highly confined. Apart from the mesh hole, it is totally surrounded by kapton on the sides and by the anode below. This feature considerably hinders the possibility of photons in the avalanche to spread and could act as a \textit{geometrical quenching}. This effect has been tentatively invoked to explain the remarkable observation that microbulk detectors present some gain even in pure noble gases~\cite{Dafni:2009jb,Tomas:2009zz,Cebrian:2010nw}. Similarly, this could be suggestive of an improved behaviour at high pressures (i.e. less degradation with pressure), something that has already been experimentally observed in Xe-TMA mixtures~\cite{Cebrian:2012sp}. The results presented in this paper qualitatively confirm this statement in Ar and Ne mixtures. We report on a very mild decrease of the maximum gain with pressure (less than a factor of 2 from 1\,bar to 10\,bar) in both the Ar and Ne mixtures tested (from 5\,bar to 10\,bar in the latter).

The work here presented has been carried out as part of the TREX-DM experiment, installed in the Canfranc Underground Laboratory (Laboratorio Subterráneo de Canfranc, LSC).  
 TREX-DM aims at the direct detection of dark matter Weakly Interacting Massive Particles (WIMPs) of particularly low masses (below $\sim$10~ GeV)~\cite{DRUKIER:2013lva}. 
The TREX-DM detector is a high-pressure TPC able to host 0.16\,kg of neon target mass at 10\,bar
(or, alternatively 0.3\,kg of argon) and built out of radiopure materials. The core of the experiment are two microbulk Micromegas readout planes with an active area of 24.8$\times 24.8$\,cm$^2$ and with an X-Y strip pattern, the largest single microbulks built so far. Amplification in gas and extreme radiopurity levels show promise to reach the combined low-threshold and low-background needed for a competitive low-mass WIMP sensitivity. The main motivation for the use of microbulk readouts in TREX-DM is their good intrinsic radioactivity, as they are very light structures composed of kapton and copper, two of the materials known to be very radiopure\,\cite{Cebrian:2010ta}. Recent measurements have set an upper limit for the $^{238}$U and $^{232}$Th natural chains of tens of\,nBq/cm$^2$\,\cite{Castel:2018gcp}. TREX-DM is operated in high pressure (currently at 4\,bar, but the pressure could be increased up to 10 bar) with Ne+2\%iC$_4$H$_{10}$ and possibly, in a different run, with Ar+1\%iC$_4$H$_{10}$\footnote{In this case, underground argon would be employed.}. The amount of quencher is set to make the mixtures non-flammable, a fact that facilitates an easier implementation of the experiment underground. The careful characterization of microbulks in these mixtures is therefore needed to assess the performance of the detector. The good behaviour with pressure that is here reported is an additional motivation for the use of such readouts in the experiment.

These results can be of interest to other projects involving high pressure TPCs; an example is the possibility of using a high pressure TPC as a near detector in future long-baseline neutrino experiments.
In Sec.~\ref{sec:Setup} and Sec.~\ref{sec:Procedure}, the experimental setup and the procedure are described.
The results in terms of electron transmission, gain and energy resolution are presented in Sec.\ref{sec:Results},
and we finish in Sec.~\ref{sec:Conclusions} with the conclusions of this work.

\section{Experimental setup}
\label{sec:Setup}
\begin{figure}[htb!]
\centering
\includegraphics[height=50mm]{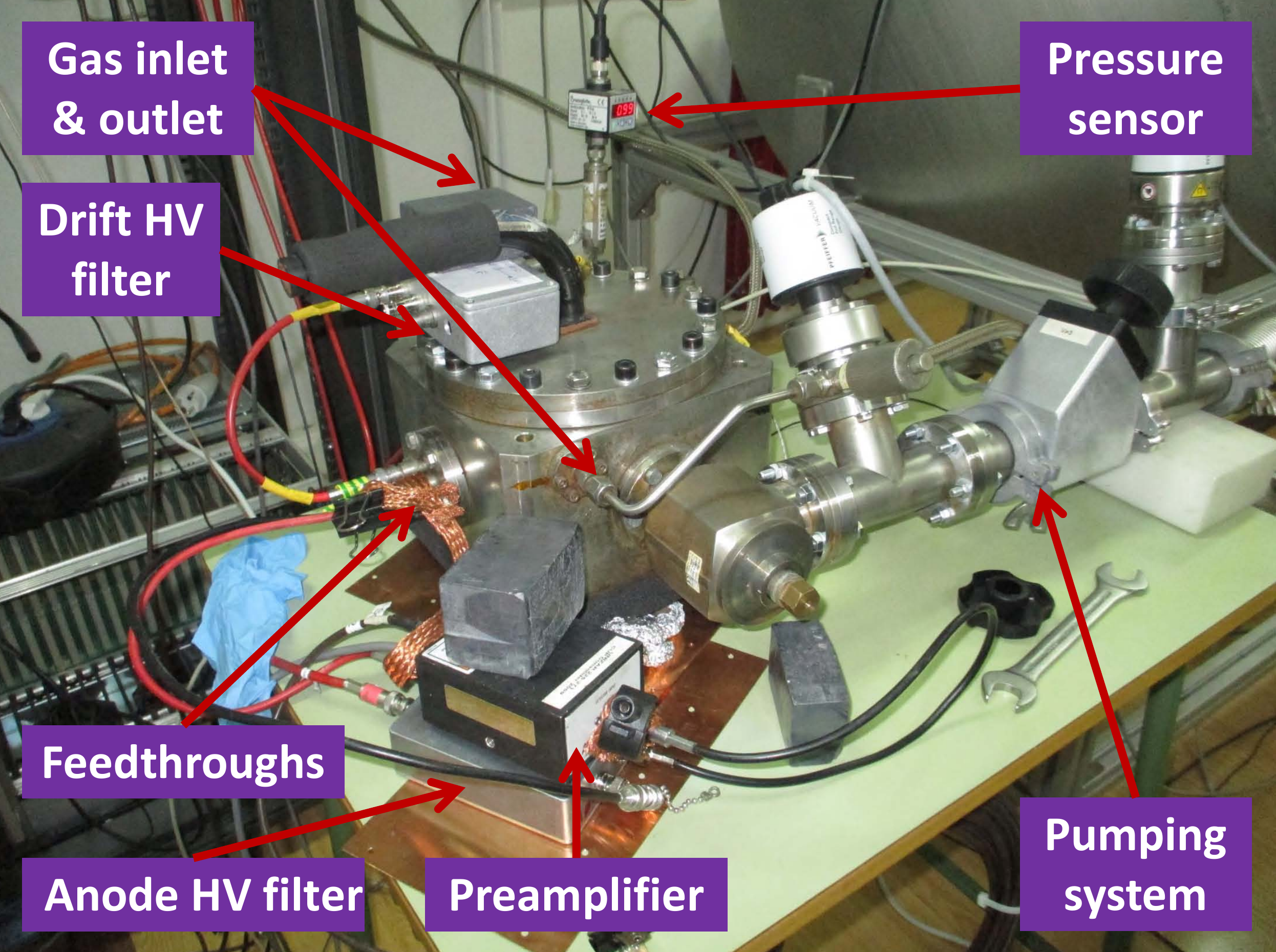}
\qquad
\includegraphics[height=50mm]{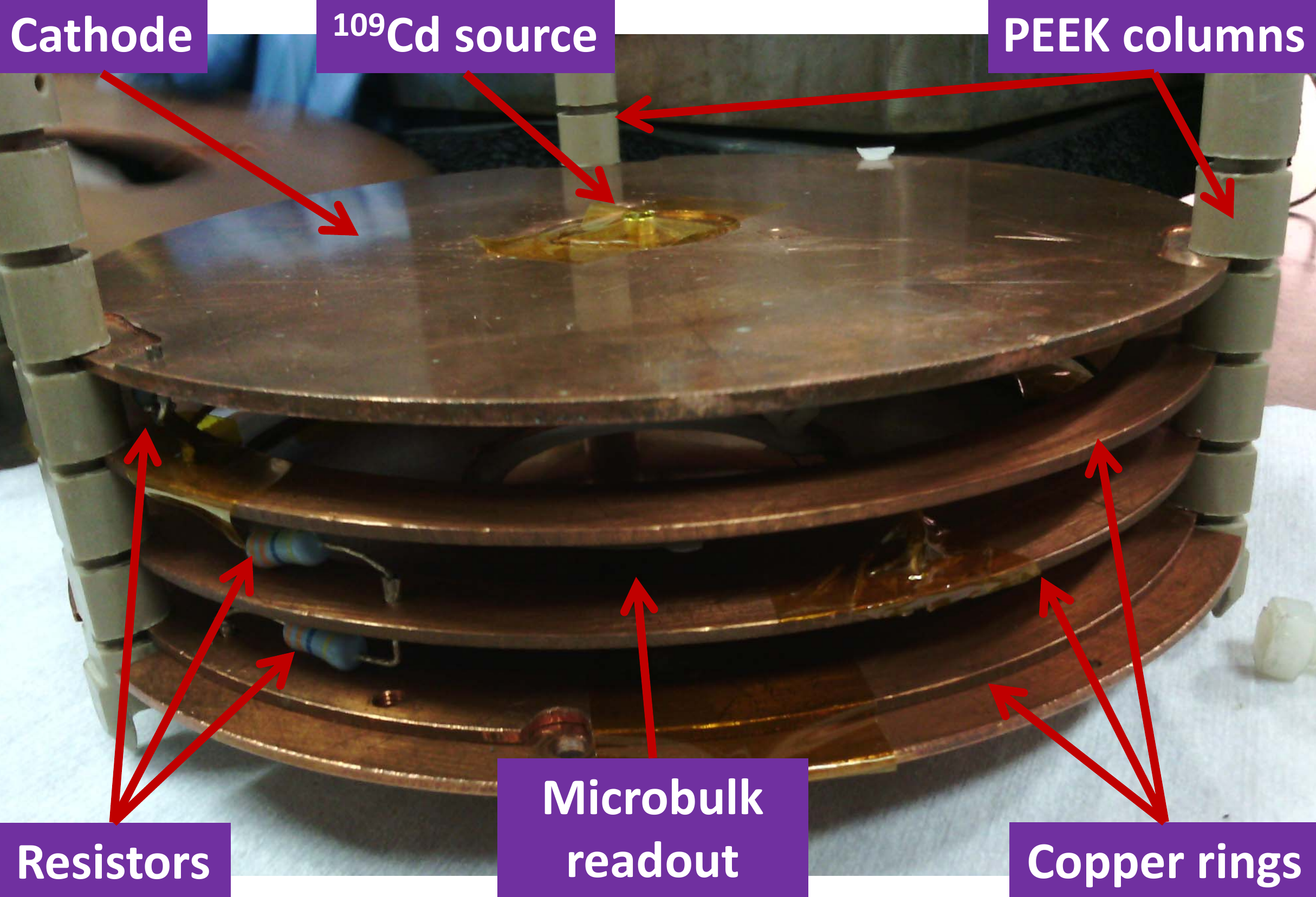}
\caption{A view of the setup (left) and the field cage (right) used in these measurements, where several items described in detail in the text are shown. Left: The chamber and the signal feedthroughs, the gas inlet/outlet,
the pressure transducer, the pumping system, the mesh and drift high voltage filters and the preamplifier. Right: The PEEK columns, the cathode, the copper rings, the resistors and the cadmium source.}
\label{fig:VesselFieldCage}
\end{figure}

The experimental setup used in these measurements (figure~\ref{fig:VesselFieldCage}, left)
is a simplified version of the one used in \cite{Cebrian:2010nw,Cebrian:2012sp}.
The main component is a stainless steel vessel\footnote{Produced by Telstar.},
which can reach vacuum values of $\sim10^{-6}$\,mbar (and outgassing rates $\sim 10^{-7}$\,mbar~l/sec)
and keep gas pressures up to 12\,bar.
Inside the vessel there is an internal cylindrical chamber (total volume of 2.2\,litres, 10\,cm height and 16\,cm diameter)
which can be accessed by a 15\,cm diameter DN160CF flange situated on the top.
At one side of the vessel, there is a DN40CF flange equipped with four SHV feedthroughs for signal extraction while at another side, there is a DN40CF outlet to pump the chamber and reduce the release of trapped air or other impurities from elements inside the vessel. This pumping system (composed of a turbomolecular pump and two vacuum gauges) is isolated from the chamber by an all-metal valve and another vacuum valve in series.
The vessel is also equipped with two DN16CF ports situated at the front (inlet) and at the back (outlet) for the gas flow. At the inlet of the vessel there is a pressure transducer\footnote{PTU-F-AC16-33AG by Swagelok.}, allowing to monitor the vessel's pressure even when it is isolated from the system via needle valves situated at the ports.

\begin{figure}[htb!]
\centering
\includegraphics[height=50mm]{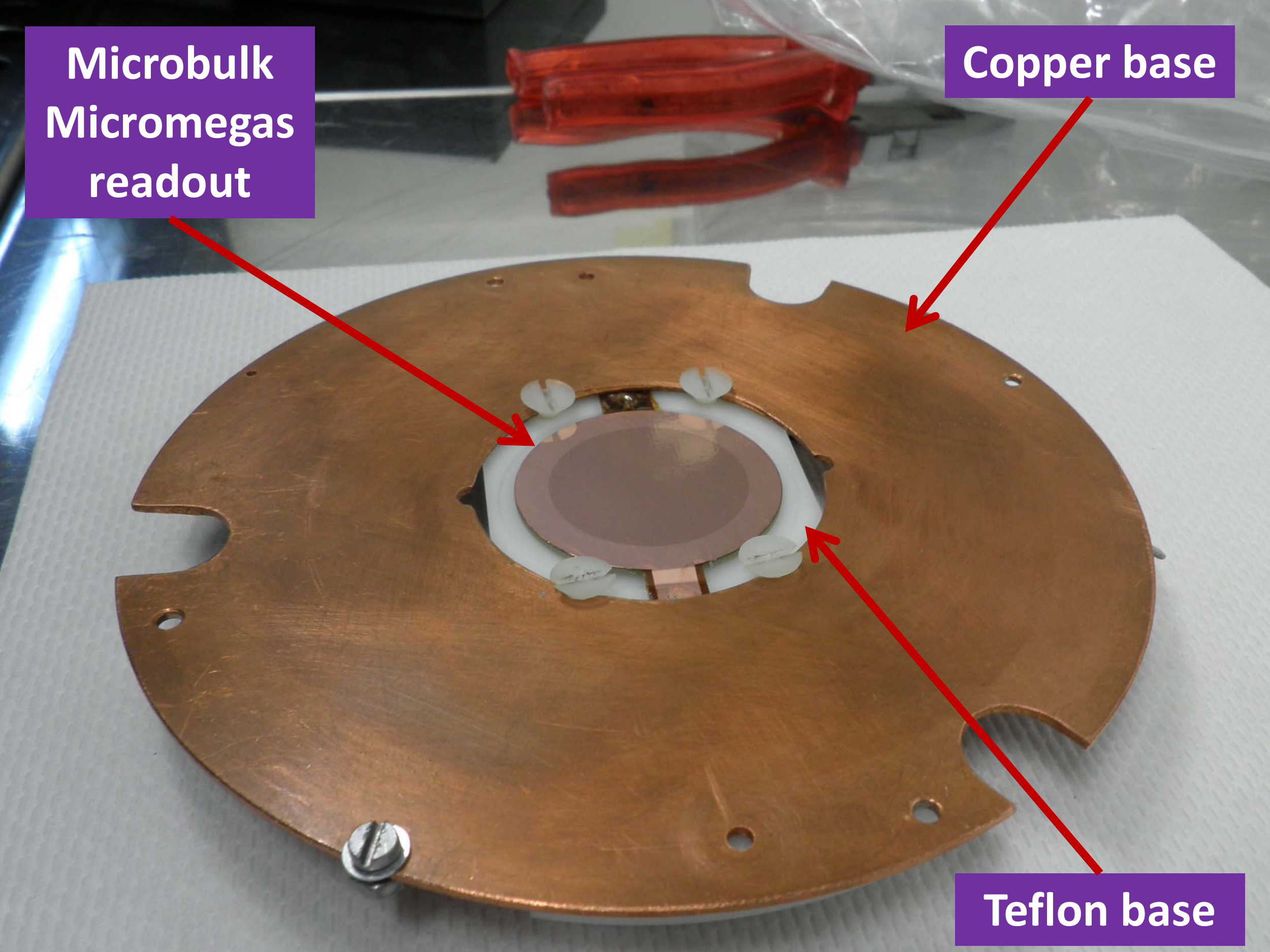}
\qquad
\includegraphics[height=50mm]{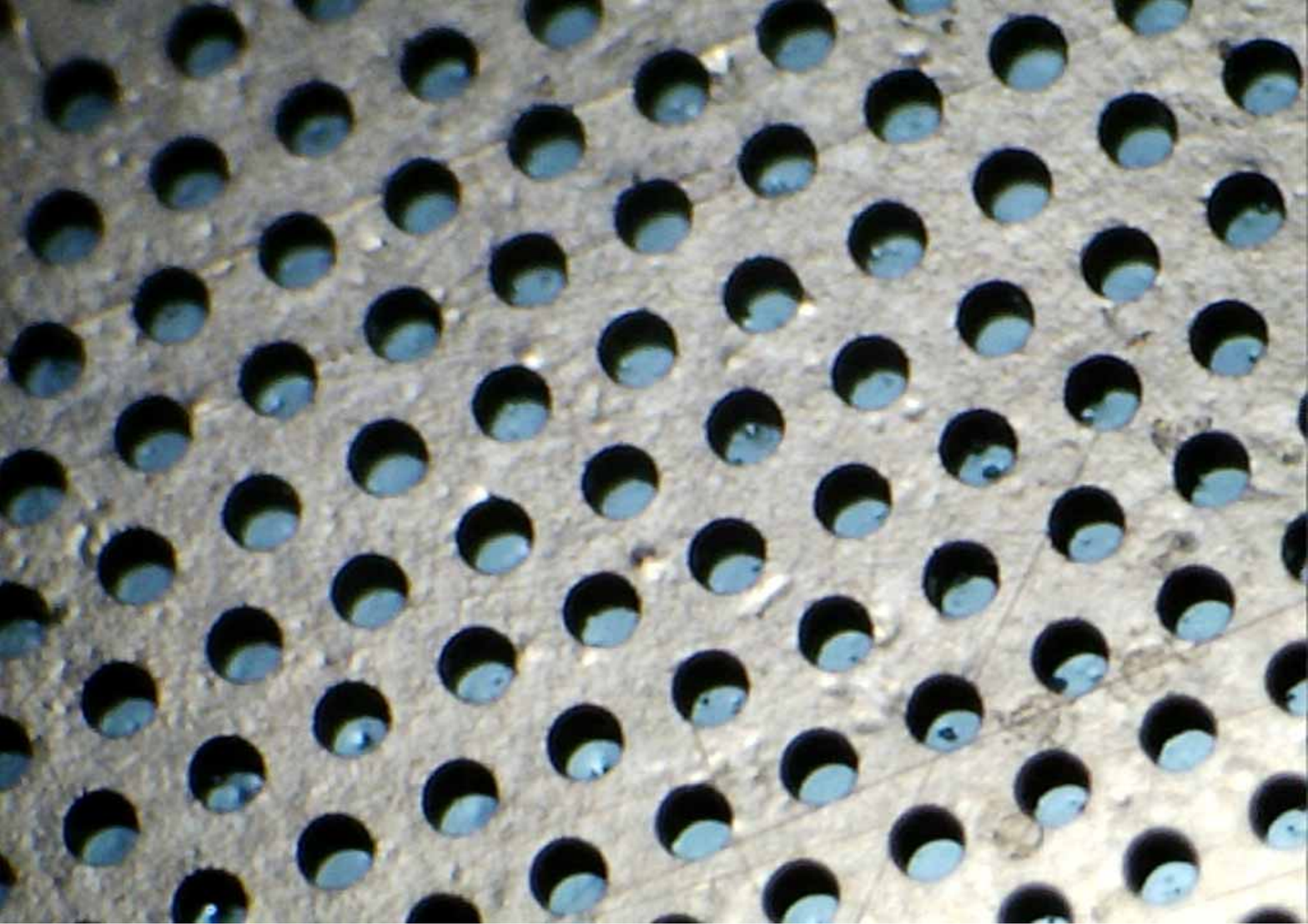}
\caption{Left: The microbulk Micromegas readout plane used in these measurements, already installed at the copper-teflon base. Right: A view at the microscope of the readout mesh. The hole pattern is hexagonal, the pitch distance is 100\,$\upmu$m and the hole diameter is 55\,$\upmu$m.}
\label{fig:MMReadoutPlane}
\end{figure}

A 3\,cm-long field cage (figure~\ref{fig:VesselFieldCage}, right) is placed in the inner chamber. This structure is composed of a 10\,cm diameter circular copper cathode and three copper rings. These elements are mechanically fixed together by three PEEK columns and are electrically linked one after the other through 10\,M$\Omega$ resistors. The cathode has a small hole at its center,
where a cylindrical aluminium container with a $^{109}$Cd deposition inside is situated.
The bottom ring sits on a circular copper plate (figure~\ref{fig:MMReadoutPlane}, left) which serves as a support for the microbulk Micromegas readout; the microbulk is fixed in a big hole at the center of the plate, isolated from the vessel by a Teflon rectangular base.
The cathode and the anode are electrically connected to the vessel feedthroughs and are independently powered by a CAEN N470A and a CAEN N471 module to a negative and a positive voltage, respectively.
The copper plate is electrically connected to the Micromegas mesh,
so as to extend the equipotential surface defined by the mesh, and set to the ground. The Micromegas readout is a single, non-segmented anode covering a 2.4\,cm diameter  area. It is denoted as G50D55P100, as it has an amplification gap of 50\,$\upmu$m and an hexagonal mesh hole pattern (figure~\ref{fig:MMReadoutPlane}, right), with a hole pitch of 100\,$\upmu$m and a hole diameter of 55\,$\upmu$m.

An event interacting in the active volume releases electrons,
which drift towards the copper plate and the Micromegas readout plane.
If these primary electrons are above the active zone of the plane,
they will pass through the Micromegas mesh holes towards the higher-field amplification region; then, they will be amplified in the gap and the charge movement will induce signals both at the mesh and the anode.
The anode signal is decoupled from the voltage level by a filter,
whose $RC$ constant minimizes the recovery time after a spark to $\sim$5\,seconds, while the mesh signal is not recorded for the measurements in this work. The anode signal is afterwards processed by a preamplifier\footnote{Model 2004 by Canberra.},
an amplifier\footnote{Model 2021 by Canberra.}
and is subsequently recorded by a Multichannel Analyzer (MCA)\footnote{Amptek MCA8000A.}.

\section{Experimental procedure}
\label{sec:Procedure}
The experimental procedure starts by pumping the vessel to pressures below $2\,\times\;10^{-5}\,$\,mbar
and outgassing rates around $9~\times~10^{-7}$\,mbar~l/sec. 
These values are considered sufficient, as no evidence of attachment in the gas has been observed in the readout performance.
The all-vacuum valve is then closed and the selected gas from a premixed bottle is injected into the vessel until a desired pressure is reached; the two gas valves are closed and the gas remains static in the vessel during the data-taking, i.e., in sealed-mode operation. After the data-taking, the gas pressure is either increased by 1\,bar to make a new measurement or the gas is removed and the vessel is pumped.

\begin{figure}[htb!]
\centering
\includegraphics[width=75mm]{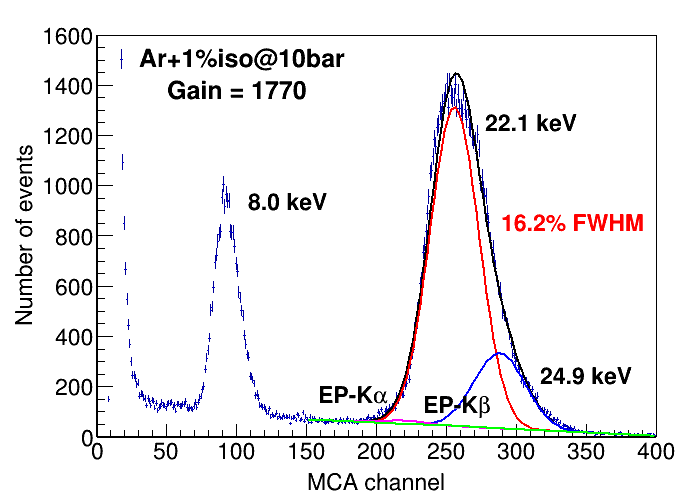}
\includegraphics[width=75mm]{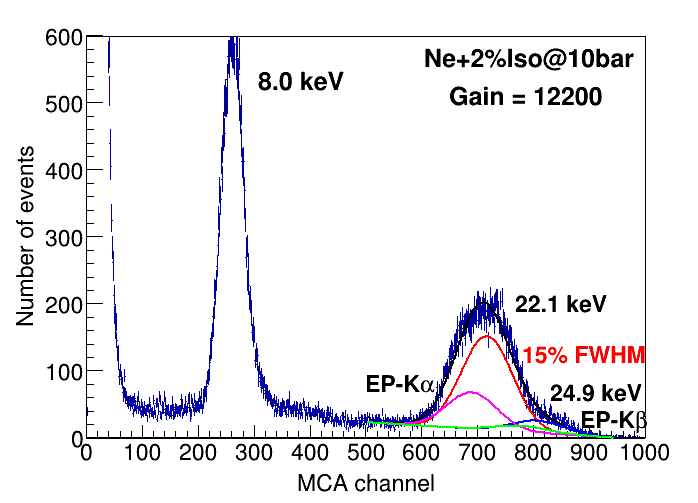}
\caption{ Energy spectra generated when the microbulk G50D55P100
was irradiated by a $^{109}$Cd source in Ar+1\%iC$_4$H$_{10}$ (left) and Ne+2\%iC$_4$H$_{10}$ (right) at 10\,bar.
The spectral parameters are calculated through an iterative multi-Gaussian fit,
which includes a linear background,
the K$_\upalpha$ ($E_\upalpha$ = 22.1\,keV, red line) and K$_\upbeta$ ($E_\upbeta$ = 24.9\,keV, blue line)
emission lines of the source
and their escape peaks located at $E_\upalpha - E_{\rm{gas}}$ (magenta line) and $E_\upbeta - E_{\rm{gas}}$ (green line).
$E_{\rm{gas}}$ is the absorption energy of the gas, which is 2.974~keV for argon and 0.849\,keV for neon.
The fluorescence lines of copper (at 8.048\,keV and 8.905\,keV) are also present.}
\label{fig:EnergySpectra}
\end{figure}

The readout has been characterized in terms of electron transmission, gain and energy resolution in Ar+1\%iC$_4$H$_{10}$ for pressures between 1\,bar and 10\,bar, in steps of 1\,bar and was calibrated with a $^{109}$Cd source. The measurements in Ne+2\%iC$_4$H$_{10}$ were performed from 5\,bar to 10\,bar in order to contain the 22\,keV energy peak of the source in the small volume of the setup.
The calibration spectra generated by the MCA (see  figure~\ref{fig:EnergySpectra}) are characterized by the K-peaks of the source and the fluorescence emissions at 8.048\,keV and 8.905\,keV
from the copper components of the vessel.
The mean position and the width of the K$_{\upalpha}$ (at $E_\upalpha$~=~22.1\,keV) is calculated
through an iterative multi-Gaussian fit, previously used in~\cite{Cebrian:2012sp}, including
both the K$_\alpha$ and K$_\upbeta$ (at $E_\upbeta$~=~24.9~keV) emission lines, their escape peaks
and a linear background. The escape peaks are located at $E_\alpha - E_{\rm{gas}}$ and $E_\upbeta - E_{\rm{gas}}$,
where $E_{\rm{gas}}$ is the absorption energy of the gas,
which is 2.974\,keV for argon and 0.849\,keV for neon.

A wide range of amplification and reduced drift fields has been scanned at each pressure.
The anode was varied, in steps of 10\,V, from 200~V to 300~V at 1\,bar
and from 460\,V to 580\,V at 10\,bar in Ar+1\%iC$_4$H$_{10}$; for Ne+2\%iC$_4$H$_{10}$ the corresponding values are from 290\,V to 470\,V at 5\,bar and from 390\,V to 560\,V at 10\,bar.
Meanwhile, the drift field was typically varied between 10\,V/cm/bar and 300\,V/cm/bar,
except for pressures near 10~bar, where the field was upper-limited to $\sim$200\,V/cm/bar; the type of SHV feedthroughs was limiting the maximum voltage applied.
These fields translate to drift voltages of 30-900\,V at 1~bar and 300-6000\,V at 10~bar.


\section{Results: electron transmission, gain and energy resolution}
\label{sec:Results}
\begin{figure}[htb!]
\centering
\includegraphics[width=75mm]{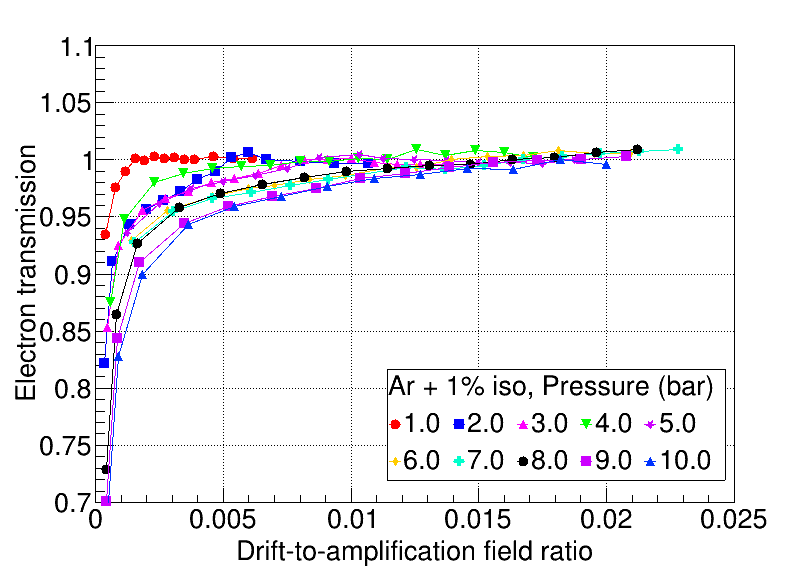}
\includegraphics[width=75mm]{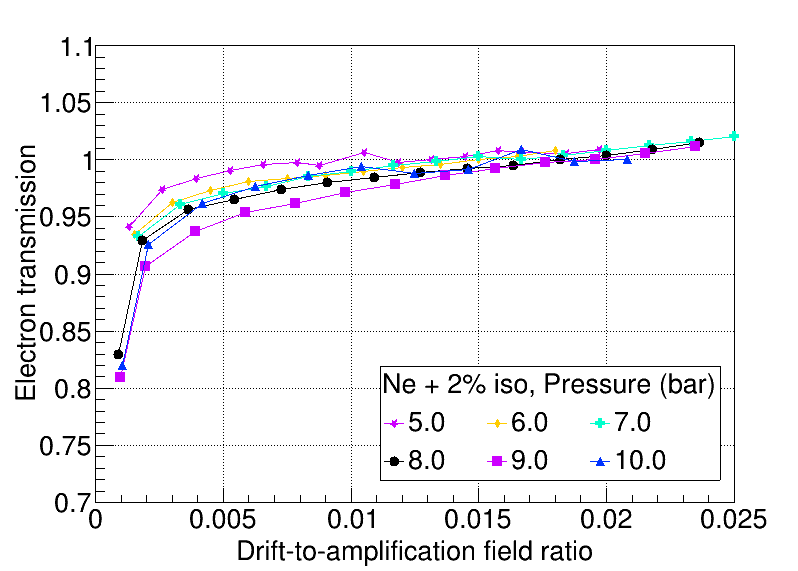}
\includegraphics[width=75mm]{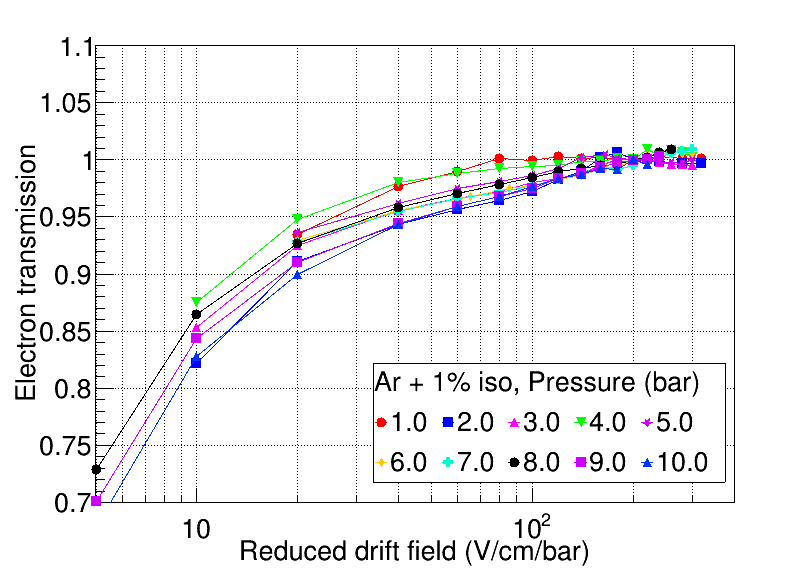}
\includegraphics[width=75mm]{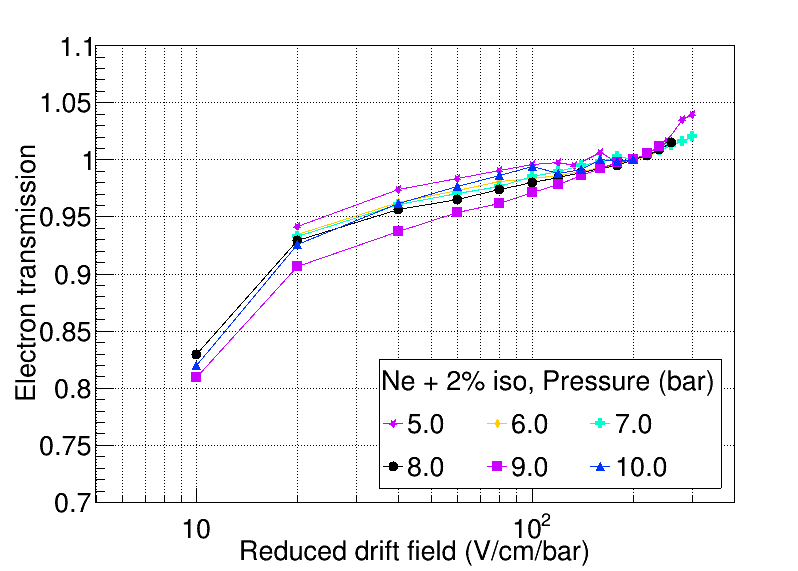}
\caption{Dependence of the electron transmission with the drift-to-amplification field ratio (top)
and the reduced drift field (bottom) in Ar+1\%iC$_4$H$_{10}$ (left, 1\,bar to 10\,bar) and Ne+2\%iC$_4$H$_{10}$ (right, 5\,bar to 10\,bar).
The gain has been normalized to the gain value at a reduced drift field of 200\,V/cm/bar.}
\label{fig:TransDriftCurves}
\end{figure}

\begin{figure}[htb!]
\centering
\includegraphics[width=75mm]{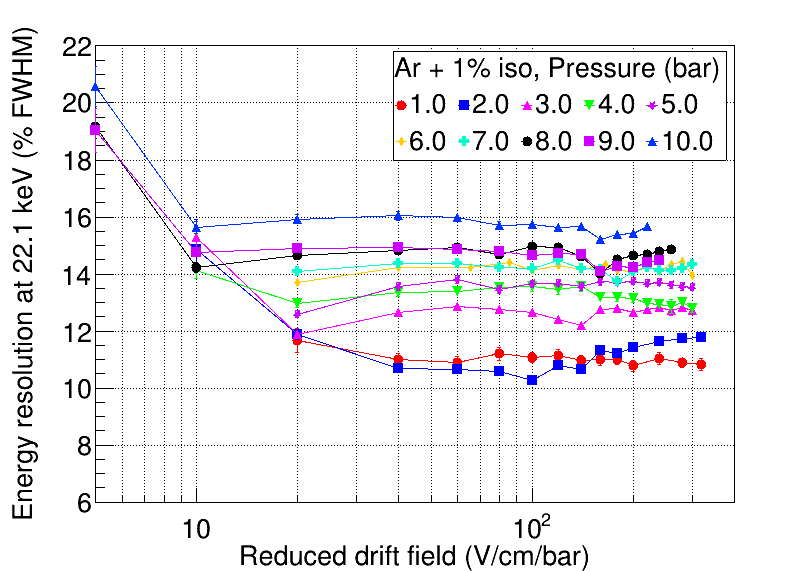}
\includegraphics[width=75mm]{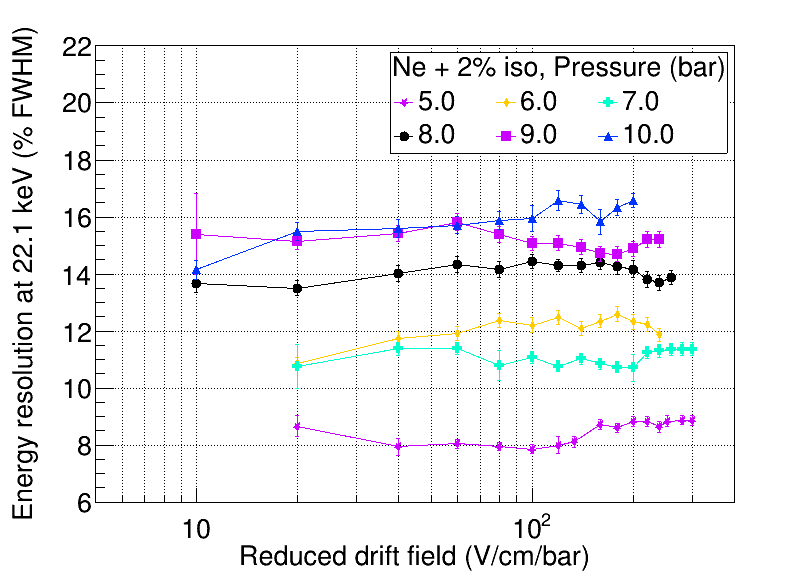}
\caption{Dependence of the energy resolution with the drift field
in Ar+1\%iC$_4$H$_{10}$ (left, 1\,bar to 10\,bar) and Ne+2\%iC$_4$H$_{10}$ (right, 5\,bar to 10\,bar).}
\label{fig:EResDriftCurves}
\end{figure}

The electron transmission is the probability for primary electrons to pass from the conversion volume to the amplification gap through the mesh holes. Its measurement includes two mechanisms that cannot be separated: the electron attachment and recombination in the drift volume, and the so-called transparency of the mesh electrode. In order to measure it, the drift voltage is varied for a fixed anode voltage to obtain the dependence of the electron transmission with the drift-to-amplification field ratio (figure~\ref{fig:TransDriftCurves}, top)
and the reduced drift field at each pressure (figure~\ref{fig:TransDriftCurves}, bottom).
The fact that the noise level during the measurements was not negligible, of the order of $\sim9 \times 10^4$ electrons, forced the starting point at detector gains of approximately 100. 
The specific anode voltages at which each electron transmission curve was taken,
as well as the gain and the energy resolution at 200~V/cm/bar are shown in Table~\ref{tab:TransValues}. As it can be seen, the detector gain was fixed between 0.4$\times 10^3$ and 1.0$\times 10^3$ for Ar+1\%iC$_4$H$_{10}$
and between 1.8$\times 10^3$ and 2.2$\times 10^3$ for Ne+2\%iC$_4$H$_{10}$.
Although this is not an absolute measurement of the electron transmission,
the fact that the signal height and the energy resolution (figure~\ref{fig:EResDriftCurves})
become independent of the drift field suggests that the mesh transmission is close to 100\% in the plateau range
and that the attachment and recombination contributions are negligible,
allowing to normalize the maximum value to the signal height.
In these measurements, we have normalized all values to the one measured at 200\,V/cm/bar; the gain variations along time have also been corrected.

\begin{table}
\centering
\caption{Voltages at which each electron transmission curve was taken, the gain and the energy resolution (\% FWHM at 22.1~keV) at 200\,V/cm/bar.
The voltages have an error of $\pm$1~V, associated to the precision of the high voltage supply. The statistical error of the gain values, derived from the fit error, is less than 0.2\%, while the one of the energy resolution is less than 0.2\% FWHM in the  argon mixture and 0.3\% FWHM in the neon one.}

\begin{tabular}{|c|cccccccccc|}
\hline
\multicolumn{11}{|c|}{Argon+1\%iC$_4$H$_{10}$}\\
\hline\hline
 P(bar) & 1.0 & 2.0 & 3.0 & 4.0 & 5.0 & 6.0 & 7.0 & 8.0 & 9.0 & 10.0\\
\hline
Anode(V) & 260 & 300 & 330 & 350 & 400 & 430 & 460 & 490 & 520 & 550\\
  G($10^2$) & 4.76 & 6.43 & 5.74 & 4.16 & 8.64 & 8.07 & 8.34 & 8.84 & 9.39 & 10.00\\
 Res. & 10.8 & 11.4 & 12.7 & 13.2 & 13.8 & 14.1 & 14.1 & 14.6 & 14.2 & 15.4\\
\hline\hline
\multicolumn{11}{|c|}{Neon+2\%iC$_4$H$_{10}$}\\
\hline\hline
 P(bar) & 1.0 & 2.0 & 3.0 & 4.0 & 5.0 & 6.0 & 7.0 & 8.0 & 9.0 & 10.0\\
\hline
Anode(V) & - & - & - & - & 380 & 400 & 420 & 440 & 460 & 480\\
G($10^3$) & - & - & - & - & 1.94 & 1.83 & 2.19 & 2.04 & 2.03 & 1.84\\
Res. & - & - & - & - & 8.8 & 12.3 & 10.7 & 14.2 & 14.9 & 16.6\\
\hline
\end{tabular}
\label{tab:TransValues}
\end{table}

The full transmission plateau is in all cases almost flat for drift fields between 100\,V/cm/bar and 300\,V/cm/bar,
with a maximum variation of 3\% for some pressures.
At very low drift fields (typically $\sim 10-20$~V/cm/bar) the electron transmission is reduced by two effects related to low drift velocities:
the charge collection time is longer than the amplifier integration time (4\,$\upmu$s) and
electron attachment and recombination of primary electrons generated in the drift volume are more important.
This \textit{drift field threshold} is similar to the one reported in~\cite{Iguaz:2012ur}
and can be explained by the short drift distance in our setup and the combination of a high gas quality and a low outgassing rate.
At high fields, the configuration of the field lines makes that some electrons get trapped in the mesh electrode.
This effect causes a drop in the mesh transmission and degrades the energy resolution.
However, neither an electron transmission drop nor a degradation in energy resolution has been observed in our measurements.
According to~\cite{Iguaz:2012ur}, the full transmission plateau should range for fields up to 150\,V/cm
for both gases at 1~bar and the detector gain should respectively drop a~20\% and 13\% when the drift field is increased
to 300\,V/cm. We have attributed the absence of this effect to the wider mesh holes of our readout,
whose diameter is 55\,$\upmu$m in comparison to the 35\,$\upmu$m of the reference.

After the measurement of the electron transmission, the drift field is set to 200\,V/cm/bar, where a full mesh transparency is systematically achieved.
The anode voltage is then varied from very low amplification fields,
where signals are just at the noise threshold, up to the spark limit,
where micro-discharges between the mesh and the anode plane produce a current higher than 300\,nA, resulting in a trip of the high-voltage power supply. After a trip, the working voltage is normally recovered within a few seconds but as a fixed criterium, the data-taking is concluded if the trip rate exceeds one every 30\,sec.
The position of the 22.1\,keV peak at the MCA is used to calculate the gain of the Micromegas readout plane, defined as the ratio of the number of electrons after the avalanche ($n$) and the number of primary electrons ($n_0$).
To obtain $n$ for each peak position, the electronic chain was previously calibrated
with a squared-pulse generator and a capacitor. The number of primary electrons $n_0$ is the ratio of the peak energy (22.1\,keV)
and the mean energy necessary for creating an electron-ion pair in the gas, which is 26.3\,eV for argon and 36.4\,eV for neon~\cite{Christophorou:1971}.
The contribution due to isobutane has been disregarded because of its low concentration.

\begin{figure}[htb!]
\centering
\includegraphics[width=75mm]{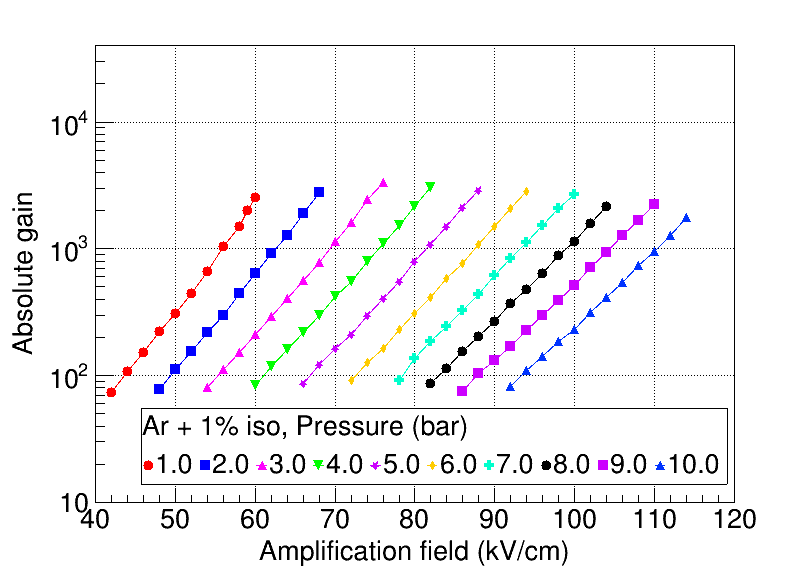}
\includegraphics[width=75mm]{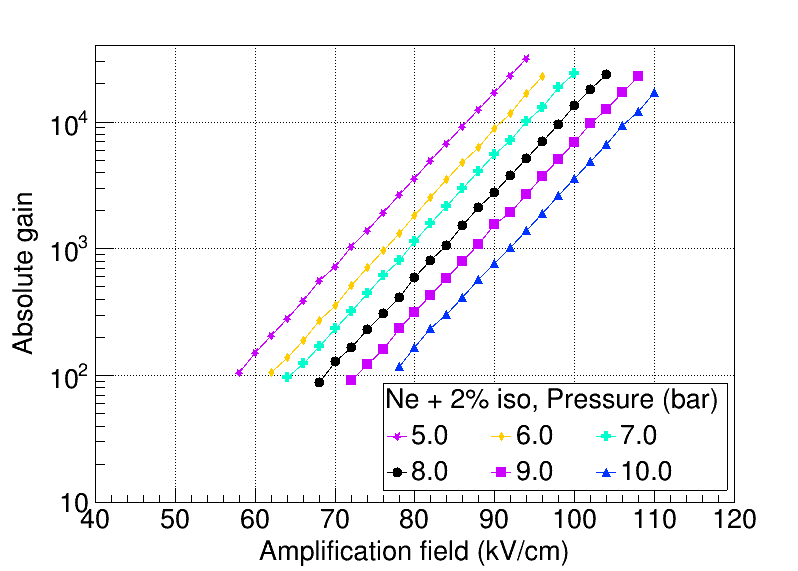}
\caption{Dependence of the absolute gain with the amplification field
in Ar+1\%iC$_4$H$_{10}$ (left, 1\,bar to 10\,bar) and Ne+2\%iC$_4$H$_{10}$ (right, 5\,bar to 10\,bar). The maximum gain of each curve is obtained just before the spark limit.
There is an uncertainty of $\pm$1\,V in the high-voltage power supply,
which produces a systematic error of $\sim$0.5\% in the determination of the amplification field.}
\label{fig:GainCurves}
\end{figure}

The gain curves obtained in Ar+1\%iC$_4$H$_{10}$ for pressures between 1~bar and 10~bar and in Ne+2\%iC$_4$H$_{10}$ between 5\,bar and 10\,bar are shown in figure~\ref{fig:GainCurves}. The readout shows a maximum attainable gain before the spark limit
higher than $1.7\times 10^3$ in the argon mixture and higher than $1.7\times 10^4$ in the neon mixture for all pressures.  The uncertainty of this point is of the order of 15\%, owing to the 10~V-step in voltage of the data-taking procedure. 
The maximum gain hardly decreases with pressure: in Ar+1\%iC$_4$H$_{10}$ from $3.4\times 10^3$ at 3\,bar
down to $1.7\times 10^3$ at 10~bar; while in Ne+2\%iC$_4$H$_{10}$ it drops from $3.1\times 10^4$ at 5\,bar
down to $1.7\times 10^4$ at 10\,bar. 
This reduction is lower than the one observed in Xe-TMA mixtures for microbulk technology~\cite{Cebrian:2012sp},
where the maximum gain dropped from $4\times 10^3$ at 1\,bar to $4\times 10^2$ at 10\,bar;
and in Ar+2\%iC$_4$H$_{10}$ for bulk technology~\cite{Iguaz:2015myh},
where the maximum gain reduced from $3\times 10^3$ at 1.2\,bar
down to $5\times 10^2$ at 10\,bar. And it is even lower than the dependence reported in~\cite{Bondar:2001mt} for a triple-GEM gaseous detector in argon,
where the maximum gain reduced from $10^5$ at 1\,bar down to $5\times 10^2$ at 5\,bar.

\begin{figure}[htb!]
\centering
\includegraphics[width=75mm]{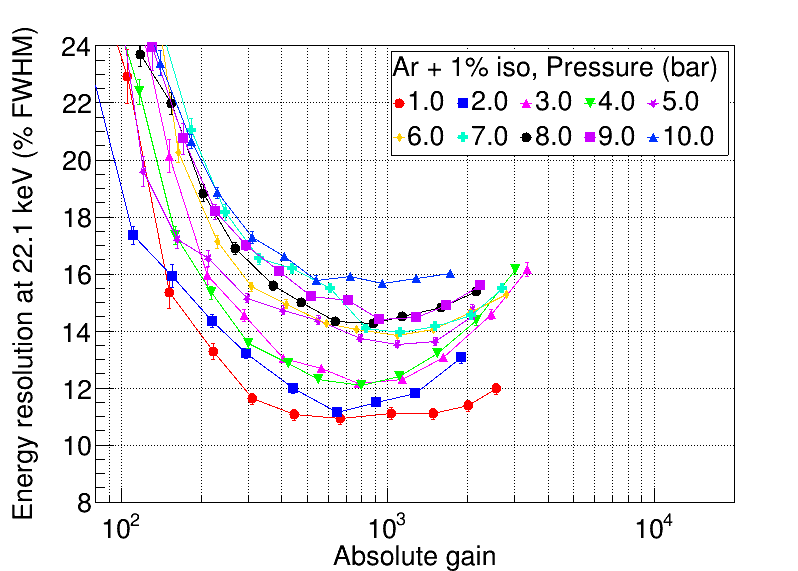}
\includegraphics[width=75mm]{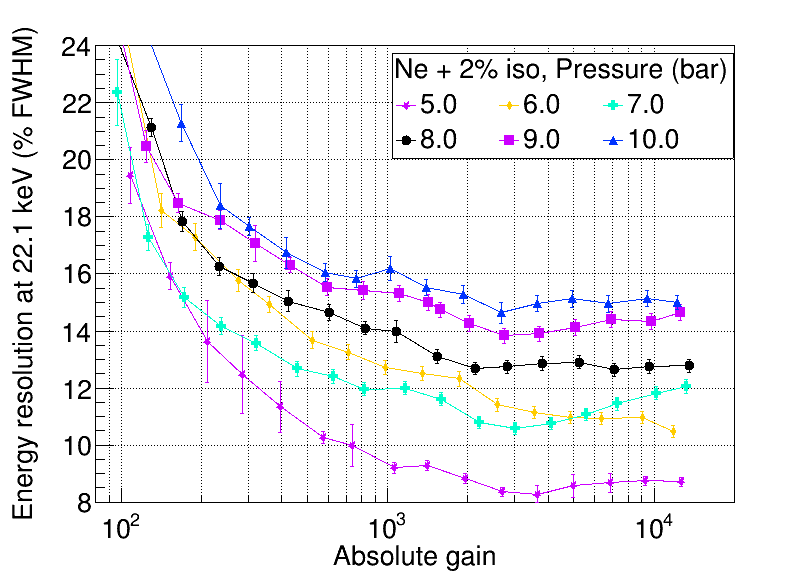}
\caption{Dependence of the energy resolution at 22.1~keV with the absolute gain
in Ar+1\%iC$_4$H$_{10}$ (left, 1\,bar to 10\,bar) and Ne+2\%iC$_4$H$_{10}$ (right, 5\,bar to 10\,bar). In the case of Ar+1\%iC$_4$H$_{10}$, the maximum gain of each curve is obtained just before the spark limit.}
\label{fig:EResGainCurves}
\end{figure}

The energy resolution (expressed in FWHM) of the readout plane is obtained from the width of the gaussian fit. Its dependence with the absolute gain is shown in figure~\ref{fig:EResGainCurves} for all pressures.
The statistical error of the energy resolution, derived from the errors from the gaussian fit parameters, is less than 0.2\% FWHM in Ar+1\%iC$_4$H$_{10}$ and 0.3\% FWHM in Ne+2\%iC$_4$H$_{10}$ for all gains,
except for low gains where it exponentially increases.
At each pressure there is a range of gains for which the energy resolution is optimized.
At low gains, the energy resolution degrades because the signal becomes comparable to the electronic noise. At high gains, the resolution degrades due to the increase in the gain fluctuations by the UV photons generated in the avalanche~\cite{Schindler201078}.

As it is shown in figure~\ref{fig:EResGainCurves}, the energy resolution at 22.1\,keV is degraded with pressure, being 10.8\%~FWHM at 1\,bar and 15.6\%~FWHM at 10\,bar for the argon mixture.
For the neon mixture, the energy resolution at 22.1\,keV worsens from 8.3\%~FWHM at 5\,bar to 15.0\%~FWHM at 10 bar.
All these values are better than those measured with the bulk readout of TREX-DM in~\cite{Iguaz:2015myh}, where a degradation from 16\% FWHM at 22.1~keV at 1.2\,bar to 25\%~FWHM at 10\,bar was observed.
Nevertheless, we do not discard that the energy resolution at low pressures may be limited by the low quantity of quencher, as better values have been measured for microbulk technology in~\cite{Iguaz:2012ur}: 11.7\% and 10.5\%~FWHM at 5.9\,keV in Ar+5\%iC$_4$H$_{10}$ and Ne+5\%iC$_4$H$_{10}$, respectively, which translates to 6.0\% and 5.4\%~FWHM at 22.1\,keV if we suppose only an energy dependence.

\section{Conclusions and prospects}
\label{sec:Conclusions}
As part of the commissioning of the TREX-DM experiment, the performance of a non-pixelated microbulk Micromegas readout has been characterized in two non-flammable mixtures: Ar+1\%iC$_4$H$_{10}$ and Ne+2\%iC$_4$H$_{10}$, for pressures between 1\,bar (5\,bar for Ne) and 10\,bar, in steps of 1\,bar.
The readout shows a maximum attainable gain before the spark limit
higher than $1.7\times 10^3$ in argon and in the case of  neon higher than $1.7\times 10^4$ for all pressures.
The maximum gain decreases with pressure by only a factor 2: in Ar+1\%iC$_4$H$_{10}$ from $3.4\times 10^3$ at 3\,bar down to $1.7\times 10^3$ at 10\,bar; while in Ne+2\%iC$_4$H$_{10}$ it drops from $3.1 \times 10^4$ at 5\,bar
down to $1.7\times 10^4$ at 10\,bar.
The energy resolution at 22.1\,keV also shows a degradation with pressure:
from 10.8\%~FWHM at 1\,bar to 15.6\%~FWHM at 10\,bar in Ar+1\%iC$_4$H$_{10}$;
and from 8.3\%~FWHM at 5\,bar to 15.0\%~FWHM at 10\,bar in Ne+2\%iC$_4$H$_{10}$. 
These results satisfy the criteria set by the TREX-DM experiment for gain and energy resolution, and might be of interest for other applications where high pressure TPC are involved. 


\acknowledgments
We thank Rui de Oliveira and his team at the CERN workshop for the microbulk readout planes
used in this work, the Servicio General de Apoyo a la Investigaci\'on-SAI of the University of Zaragoza for the fabrication of some mechanical components and our IRFU/CEA-Saclay colleagues for their always wise advice.
We acknowledge the support from the European Research Council under the European Union's IDEAS program of the 7th EU Framework Program and under the Horizon 2020 research and innovation programme, grant agreements ERC-2009-StG-240054 (T-REX project) and ERC-2017-AdG-788781 (IAXO+ project) 
and the Spanish Agencia Estatal de Investigación (AEI) under grants FPA2016-76978-C3-1-P and the coordinated grant PID2019-108122GB. T.D acknowledges the support from the \emph{Ram\'on y Cajal} program of the MINECO.









\bibliographystyle{JHEP}
\bibliography{20201227_MMArNeIso}
\end{document}